# General Relativity from Intersection Theory

Hjalte Frellesvig[1,2,*] and Toni Teschke[2,†]

[1]*Niels Bohr International Academy, University of Copenhagen, Blegdamsvej 17, 2100 Copenhagen, Denmark*
[2]*Niels Bohr Institute, University of Copenhagen, Blegdamsvej 17, 2100 Copenhagen, Denmark*

This paper combines the post-Minkowskian expansion of general relativity with the language of intersection theory. Because of the nature of the soft limit inherent to the post-Minkowskian expansion, the intersection-based approach is of enhanced utility in that theory compared to a generic quantum field theory. In the language of intersection theory, Feynman integrals are rephrased in terms of twisted cocycles. The intersection number is a pairing between two such cocycles and its existence allows for the direct projection onto a basis of master integrals. In this paper we use this approach to compute the second post-Minkowskian contribution to the scattering of two compact astronomical objects, getting results in agreement with previous findings.

## I. INTRODUCTION

The groundbreaking discovery of gravitational waves [1, 2] from colliding compact astronomical objects, such as black holes and neutron stars, led to much renewed interest in methods for analytical precision calculations in the theory of general relativity. One such method is the *post-Minkowskian expansion* (PM) [3, 4] which treats the problem perturbatively in Newton's constant $G_N$, but keeps it exact in all other quantities. This expansion is reminiscent of perturbative quantum field theory (QFT), and indeed it turns out that many aspects of QFTs can be adapted to this completely classical problem [5–12] (see e.g. refs. [4, 13] for summaries). In particular, the inspiraling phase that is of the highest astrophysical interest may be mapped to the scattering problem often studied in perturbative QFT through an analytical continuation [14]. Furthermore, in this paper we will only consider the *scalar approximation* in which spin effects and internal structure of the astronomical objects are disregarded, allowing us to treat them as scalars in the QFT context [6].

When performing precision calculations in a QFT, it is usually the computation of the *Feynman integrals* that is the source of the most inconvenience and the cause of the greatest computational effort. It is therefore worthwhile to make sure that the number of Feynman integrals that has to be computed is minimal. A way to do this is to express the Feynman integrals appearing in the problem in terms of a minimal set of objects, known as *master integrals* which form a basis for the vector space of Feynman integrals of a given scattering amplitude. A reduction to a basis of master integrals is traditionally done using integration by parts (IBP) identities [15] as systematized by Laporta's algorithm [16] and implemented in a number of public [17, 18] computer programs. Laporta's algorithm has as a necessary step the solution of a system of linear equations relating various Feynman integrals, and that system can grow to enormous size for complicated Feynman integral families. That makes this step a genuine bottleneck, and it motivates the search for alternative approaches to deriving Feynman integral relations.

One such approach to the master integral decomposition of Feynman integrals has been developed recently, under the headline of *intersection theory*. This theory [19–21] had its origin in the study of (Aomoto-Gel'fand) hypergeometric functions, a class of function to which Feynman integrals in dimensional regularization evaluate when treated as exactly in $d$. In a groundbreaking paper [22] it was realized that this mathematical formalism provides a natural framework in which to analyze Feynman integrals, and in particular to derive the relations between them, something that until then had been the purview of the IBP framework. This work led to many further developments and refinements [23–30] of the theory, with the complexity of the integrals that can be successfully tackled with the intersection-based approach increasing quickly. For summaries of that theory, see e.g. refs. [31, 32].

In this paper, we will combine these two approaches, the post-Minkowskian expansion and the intersection approach for Feynman integral relations. This extends the work of ref. [33] that outlined the feasibility for the first time. Combining these two approaches reveals that the structure of the integrals needed for the PM expansion makes the intersection approach more powerful in this setting than it is for a generic QFT. We limit our discussion to the second post-Minkowskian (2PM) amplitude contribution, corresponding to Feynman integrals with one loop. We recognize, of course, that this is a well-studied problem [5–7, 10] (the state of the art is 4PM [34–36] with the first published results at 5PM appearing in the literature [37–39]). Yet we will claim that this setting is sufficient to make our point and show the enhanced applicability of the intersection-based approach to Feynman integrals in the PM expansion.

## II. CONVENTIONS AND SETUP

We will consider the scattering process as depicted in Fig. 1. The momenta are defined as "all incoming" so

[*] hjalte.frellesvig@nbi.ku.dk
[†] teschke.toni12@gmail.com

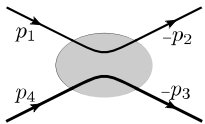

FIG. 1: The kinematics considered. The arrows indicate whether the scalar is incoming or outgoing.

$p_1+p_2+p_3+p_4 = 0$. In particular, we have

$$p_1^2 = p_2^2 = m_1^2, \qquad p_3^2 = p_4^2 = m_2^2 \qquad (1)$$

Furthermore we define the Mandelstam variables

$$s = (p_1 + p_4)^2, \qquad t = (p_1 + p_2)^2 \qquad (2)$$

with the third Mandelstam variable being related through

$$u = (p_1 + p_3)^2 = 2m_1^2 + 2m_2^2 - s - t \qquad (3)$$

To obtain the Feynman rules for a given field theory, it is usually helpful to write down its action. For a scalar gravitational interaction, the action has the following form:

$$S = \frac{2}{\kappa^2}\left(S_{\rm EH} + S_{\rm gf}\right) + S_\phi \qquad (4)$$

where $S_{\rm EH}$ is the Einstein-Hilbert action, $S_{\rm gf}$ is the gauge fixing action, $S_\phi$ is the scalar action, and $\kappa^2 = 32\pi G_N$. To perform the perturbative expansion we linearize the metric $g_{\mu\nu} = \eta_{\mu\nu} + h_{\mu\nu}$. For the gauge fixing, we chose the *de Donder gauge*. That gauge choice is defined by $\eta^{\mu\nu}\partial_\mu g_{\sigma\nu} - \frac{1}{2}\eta^{\mu\nu}\partial_\sigma g_{\mu\nu} = 0$ and is thus linear in $h_{\mu\nu}$ making it suitable for our purposes. For the derivation and the specific expressions for the Feynman rules, see e.g. refs. [33, 40, 41].

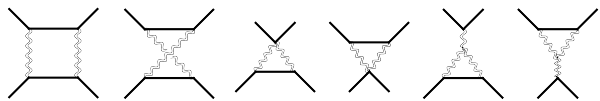

FIG. 2: The six Feynman diagrams that give a nonvanishing contribution in the soft limit.

A part of the post-Minkowskian expansion corresponds to taking the classical or *soft* limit of the momentum exchange between the two scalars [42]. With our definitions here, that corresponds to the limit $t \to 0$. Of the Feynman diagrams naively contributing to the scattering process, some will vanish in that limit. In particular there will be no contributions from diagrams containing *tadpoles* or *bubbles*. With this in mind, the diagrams that give a nonvanishing contribution the 2PM amplitude are depicted in Fig. 2.

The first diagram, the box, may schematically be written as a Feynman integral as

$$\int \frac{d^d k}{i\pi^{d/2}} \frac{N(k)}{k^2\left((k+p_1)^2-m_1^2\right)(k+p_1+p_2)^2\left((k-p_4)^2-m_2^2\right)} \qquad (5)$$

where $N(k)$ is the numerator of the Feynman diagram. Similar expressions hold for the remaining diagrams.

To apply intersection theory, it is useful to transform the Feynman integrals to a parametric representation, with the *Baikov representation* [43] being the most convenient. At one loop, the Baikov representation reads as follows:

$$I = \frac{\mathcal{J}\mathcal{G}^{\frac{E-D+1}{2}}}{\Gamma\left(\frac{D-E}{2}\right)i\pi^{\frac{E}{2}}} \int \frac{N(\mathbf{z})\mathcal{B}(\mathbf{z})^{\frac{D-E-2}{2}}d^{E+1}\mathbf{z}}{z_1^{a_1}\cdots z_n^{a_n}} \qquad (6)$$

Here $E$ is the number of independent external momenta (i.e. 3 in our case), $D$ is the spacetime dimensionality, $z_i$ are the Baikov variables that at one loop equal the propagators of the diagram in the top sector, and $a_i$ are the powers of those propagators. $\mathcal{G}$ and $\mathcal{B}$ are Gram determinants of the external momenta, and all momenta, respectively, i.e.

$$\mathcal{B} = \det(G(k,p_1,p_2,p_3)), \quad \mathcal{G} = \det(G(p_1,p_2,p_3)) \quad (7)$$

with $\mathcal{B}$ being known as the *Baikov polynomial*. Lastly, $\mathcal{J} = 2^{-E}$ is the Baikov Jacobi determinant. For more on the Baikov representation see e.g. refs. [31, 44, 45].

### III. INTERSECTION THEORY

In this section we will summarize the parts of the mathematical topic of intersection theory relevant for our current problem. We will first, in section III A, set up our notation and describe the mathematical framework. Then, in section III B we will describe how to use an inner product to perform a reduction to a master integral basis. In section III C we will then introduce that inner product as the intersection number in the univariate case. From there we will discuss the generalization to the multivariate case in section III D. Finally, in section III E, we will introduce the concepts of delta-bases and relative cohomology.

#### A. Framework: Twisted Cohomology

As master integrals form a basis for a given family of Feynman integrals, decomposing a Feynman integral into master integrals can be simplified if the space has a structure allowing for an inner product. Intersection theory suggests such a structure exists for Feynman integrals, facilitating a new method of determining the coefficients. To begin with, we will consider a general multivariate integral that has a form:

$$\int_\gamma P_1(\mathbf{z})^{\alpha_1}\cdots P_m(\mathbf{z})^{\alpha_m} d\mathbf{z} \qquad (8)$$

where $P_i(\mathbf{z})$, $i \in \{1,\ldots,m\}$ are multivariate polynomials in $\mathbf{z}$, $\alpha_i \in \mathbb{CP}^n/\mathbb{Z}$ where $\mathbb{CP}^n$ is the complex projective space, and $\gamma$ is the integration contour that lies in $X = \mathbb{C}/\bigcup_{j=1}^{m} \mathcal{D}_j$. $\mathcal{D}_j$ is a hypersurface defined by the polynomial equation $P_i(\mathbf{z}) = 0$, such that $\mathcal{D}_j = \{\mathbf{z} \in \mathbb{C} : P_j = 0\}$, and $\mathcal{D} = \bigcup_{i=1}^{m} \mathcal{D}_i$, which is termed the divisor.

$P_i(\mathbf{z})^\alpha$ can be separated into a multivalued function $u(\mathbf{z})$ on $X$ and a part belonging to the differential of $d\mathbf{z}$ that together form a smooth $n$-form $\varphi$ on $X$.

We notice that the Baikov representation of eq. (7) is of the form given by eq. (8) (at least if we take the $a_i$ of eq. (7) to be generic). Specifically, we may (ignoring constant prefactors) write the integral as

$$I = \int_\gamma u(\mathbf{z})\,\varphi(\mathbf{z}) \qquad (9)$$

where

$$u(\mathbf{z}) = \mathcal{B}^{\frac{D-E-2}{2}} \quad \text{and} \quad \varphi(\mathbf{z}) = \frac{N(\mathbf{z})}{z_1^{a_1} \cdots z_n^{a_n}} d\mathbf{z} \qquad (10)$$

To determine the integral of a multivalued $n$-form $u\varphi$ on $\gamma$ in $\mathcal{D}$, it is necessary to define a branch of $u$ on $\gamma$, that is $\gamma \otimes u_\gamma$, with $u_\gamma$ being a fixed branch of $u$ that is integrated along $\gamma$. The integral thus formally becomes $I = \int_{\gamma \otimes u_\gamma} u(\mathbf{z})\varphi(\mathbf{z})$. We will in the following simply write $u$ instead of $u_\gamma$. In the notation of intersection theory, the integral eq. (9) reads

$$\int_\gamma u(\mathbf{z})\varphi(\mathbf{z}) \;=\; \langle \varphi | \gamma \otimes u(\mathbf{z}) ] \;=\; \langle \varphi | \mathcal{C} ] \qquad (11)$$

(with the latter equation providing the definition of $|\mathcal{C}]$). $\langle \varphi |$ and $|\mathcal{C}]$ are known as a *twisted cocycle* and *cycle*, respectively, and are to be understood as members of the twisted *cohomology* group $\mathrm{H}^n$ and the twisted *homology* group $\mathrm{H}_n$ [46]. That is,

$$\langle \bullet | \bullet ] : \mathrm{H}^n(X, \nabla_\omega) \times \mathrm{H}_n(X, u) \to \mathbb{C} \qquad (12)$$

In the Baikov representation, IBP identities may be expressed as

$$0 = \int_\gamma d\left( \frac{\mathcal{B}^{\frac{D-E-2}{2}}}{z_1^{a_1} \cdots z_n^{a_n}} \zeta \right) \qquad (13)$$

where $\mathcal{B}(\partial \gamma) = 0, \zeta$ is an $(n-1)$ differential form $\zeta = \sum_{i=1}^{n} (-1)^{i+1} \zeta_i$, with $\zeta_i = \hat{\zeta} dz_1 \wedge \cdots \wedge \hat{dz_i} \wedge \cdots \wedge dz_n$ and each $\widehat{\zeta_i}$ a rational function in the variables $\mathbf{z} = (z_1, \ldots, z_n)$ whose coefficients may depend on the kinematic invariants. Here we may identify $\xi = \frac{\zeta}{z_1^{a_1} \ldots z_n^{a_n}}$, where different $\xi$ correspond to the different integrals with their respective integers that are linked through IBPs. We may thus write eq. (13) as

$$0 = \int_\gamma d(u(\mathbf{z})\xi) = \int_\gamma u(\mathbf{z}) \nabla_\omega \xi \qquad (14)$$

The covariant connection $\nabla_\omega$ on the manifold $X$ is defined as

$$\nabla_\omega \xi = d(\xi(\mathbf{z})) + \omega(\mathbf{z}) \wedge \xi(\mathbf{z}) \qquad (15)$$

where $\omega(\mathbf{z})$ is a holomorphic one-form and the connection coefficient of the manifold $X$

$$\omega(\mathbf{z}) = d\log u(\mathbf{z}) = \frac{du(\mathbf{z})}{u(\mathbf{z})} = \sum_i \hat{\omega}_i dz_i \qquad (16)$$

The covariant connection is an element of the tangential vector space that is formed by the IBPs. Eq. (14) implies that the integral is invariant under a shift in the single-valued one-form $\varphi$ by $\varphi + \nabla_\omega \xi$:

$$\int_\gamma u(z)\varphi = \int_\gamma u(z)(\varphi + \nabla_\omega \xi) \qquad (17)$$

Intuitively, $\varphi \to \varphi + \nabla_\omega \xi$ can be visualized as a gauge transformation on the tangential vector bundle of $X$. This suggests that there exists an equivalence class of $\varphi$ and $\varphi + \nabla_\omega \xi$, and indeed the cocycles $\langle \varphi |$ are representatives of these equivalence classes. These ideas were first formulated in the work of refs. [23, 31, 32, 47].

As discussed in the introduction, the motivation for considering the intersection number is that it may play the role of an inner product between Feynman integrals. What it is in detail is an inner product, not between the Feynman integrals themselves, but between the cocycles and their *duals*

$$\langle \varphi | \varphi^\vee \rangle \qquad (18)$$

where the dual cohomology group is defined by integrals of the form of eq. (9), except that $u \to u^{-1}$, i.e. $I^\vee = \int_\gamma u^{-1} \phi^\vee$. For a discussion of the physical interpretations of these dual integrals, see e.g. refs. [28, 48, 49]. Before giving any explicit definitions, let us show in generality how such an inner product can help with performing reductions onto a master integral basis.

### B. Integral Reductions Using Projections

We may write a given Feynman integral in terms of a basis of master integrals, as

$$I = \sum_{i=1}^{\nu} c_i \mathcal{I}_i \qquad (19)$$

where $\nu$ is the number of master integrals in the given family. The integrals on the right-hand side (RHS) are the master integrals, and they may be written as

$$\mathcal{I}_i = \int_\gamma u(\mathbf{z}) e_i(\mathbf{z}), \qquad (20)$$

where, importantly, $u$ and $\gamma$ are the same as in eq. (9) and the different integrals differ only by the value of the $e_i$.

Furthermore, we have that

$$\nu \;=\; \#\text{ of solutions of ``}\omega = 0\text{''} \qquad (21)$$

with the $\omega$ being given by eq. (16). This is known as the Lee-Pomeransky criterion after ref. [50] and is equivalent to the statement $\nu = \dim(\mathrm{H}^n) = \dim(\mathrm{H}_n)$.

The decomposition of an arbitrary Feynman integral of the form eq. (9) into its master integral basis corresponds, in the language of intersection theory, to the decomposition of $\langle\varphi| \in \mathrm{H}^n(X,\nabla_\omega)$ into the basis $\langle e_1|,\ldots,\langle e_\nu| \in \mathrm{H}^n(X,\nabla_\omega)$. A similar decomposition may be performed for their respective duals. It follows that

$$I = \langle\varphi|\mathcal{C}] = \sum_{i=1}^{\nu} c_i \langle e_i|\mathcal{C}] \quad (22)$$

This is master integral decomposition in the language of intersection theory.

We are now ready to express the decomposition coefficients $c_i$ in terms of our inner product: We may formally define such an inner product between our basis cocycles $\langle e_i|$ and a corresponding set of dual cocycles $|h_j\rangle$. In fact, we may define a Gram matrix for these cocycles:

$$C_{ij} = \langle e_i|h_j\rangle \quad (23)$$

This matrix will have dimensions $\nu \times \nu$ and be of full rank. We may express unity as $\mathbb{I}_c = \sum_{i,j=1}^{\nu} |h_j\rangle (\mathbf{C})^{-1}_{ij} \langle e_j|$. Multiplying on the left with $\langle\varphi|$ and comparing with eq. (22) yields

$$\langle\varphi| = \sum_{i=1}^{\nu} c_i \langle e_i| \quad \text{with} \quad c_i = \sum_{j=1}^{\nu} \langle\varphi \mid h_j\rangle \left(\mathbf{C}^{-1}\right)_{ji} \quad (24)$$

This is the *master decomposition formula*. Importantly, the values of $c_i$ are the same as in eq. (22); the choice of contour plays no role. A detailed derivation of eq. (24) can, for example, be found in refs. [22, 32, 33]. Thus, we realize why a consistent definition of an inner product between our twisted cocycles may be of great help for the very practical purpose of master integral decomposition.

### C. The Univariate Intersection Number

Let us begin by discussing how to define an intersection number in the univariate case, i.e. the case where the integrals of eq. (9) involve one integration only. This derivation comes from ref. [20], and see also refs. [22, 31, 32, 51] for further discussion.

The intersection number is a bilinear, nondegenerate pairing among the elements $\langle\varphi| \in \mathrm{H}^1(X,\nabla_\omega)$ and their dual $|\varphi^\vee\rangle$.

Following our intuition that the intersection number defines an inner product, it takes the form

$$\langle\varphi|\varphi^\vee\rangle = \int_X \mathcal{R}(\varphi(z)) \wedge \varphi^\vee(z) \quad (25)$$

with $\langle\bullet \mid \bullet\rangle : \mathrm{H}^1(X,\nabla_\omega) \times \mathrm{H}^1(X,\nabla_{-\omega}) \to \mathbb{C}$. The operator $\mathcal{R}_\omega(\varphi)$ is a regulator that is needed to account for the potentially divergent behaviour at the edge of $X$ and the fact that the numerator of the integral may have a $dz \wedge dz = 0$. In particular, problems may arise near the members of $\mathcal{P}$, which consists of the set of poles of $\omega = d\log|u|$, i.e. the set of critical points of $\log|u(z)|$ including $\infty$.

Since $\varphi$ and $\mathcal{R}_\omega(\varphi)$ are in the same cohomology class, they only differ by a covariant derivative:

$$\varphi - \mathcal{R}_\omega(\varphi) = \nabla_\omega \xi \quad (26)$$

$\mathcal{R}$ is defined such that $\xi$ is given by

$$\xi = \sum_{i \in \mathcal{P}} h_i \psi_i \quad (27)$$

where $h_i$ is the following bump-shaped function:

$$h_i(z,\bar{z}) = \begin{cases} 1 & \text{on } U_i \\ 0 \leq h_i \leq 1 & \text{on } V_i \backslash U_i \\ 0 & \text{outside } V_i \end{cases} \quad (28)$$

where the regions $U_i$ and $V_i$ are centered on the corresponding member of $\mathcal{P}$ (i.e. a pole of $\omega$). See e.g. refs. [21, 32] for more discussion along with explanatory figures.

On $U_i/z_i$, where $\mathcal{R}(\varphi) = 0$, we have the following differential equation:

$$\nabla_\omega \psi_j = \varphi \quad (29)$$

If in the integral eq. (25) we rewrite $\mathcal{R}(\varphi)$ using eqs. (26) and (27), write the covariant connection on the manifold as eq. (15), and consider the cases given in eq. (28), we arrive after a short calculation at the following expression:

$$\langle\varphi|\varphi^\vee\rangle = -\sum_{i=1}^{m} \int_{\partial V_i} \psi_i \varphi \quad (30)$$

Using the residue theorem, we may finally derive

$$\langle\varphi \mid \varphi^\vee\rangle = -2\pi i \sum_{x_i \in \mathcal{P}_\omega} \mathrm{Res}\left(\psi_i \varphi^\vee\right) \quad (31)$$

This is the equation we will use for the computation of intersection numbers in the univariate case, though for practical purposes we will suppress the $2\pi i$ prefactor.

Using eq. (31) requires the solution of the differential equation eq. (29). This may be done using a power series expansion, in which the coefficients may be found recursively [32, 52]. If we let $f_i$ denote the $i$th term in the series expansion of $f$, and let $\min_f$ denote the smallest $i$ for which $f_i$ is nonzero, the recursive solution is

$$\psi_{\min_\varphi + 1} = \frac{\hat{\varphi}_{\min_\varphi}}{\min_\varphi + 1 + \hat{\omega}_{-1}} \quad (32)$$

$$\psi_{m+1} = \frac{1}{m + 1 + \hat{\omega}_{-1}} \left(\hat{\varphi}_m - \sum_{q=0}^{m-\min_\varphi - 1} \hat{\omega}_q \psi_{m-q}\right)$$

where the coefficients on the RHS are known up to the $m$th order. Although it is possible to calculate the coefficients up to an arbitrary order, our primary interest is in the lower order terms, since the objective is to determine the residue in eq. (31).

### D. The Multivariate Intersection Number

We are now ready to discuss the intersection number in the more involved multivariate case. The multivariate intersection number was first introduced in ref. [51] and further discussed in refs. [23, 24]. The multivariate intersection algorithm discussed here has been successfully applied in the context of Feynman integrals as well as for hypergeometric functions [23, 24]. Multivariate intersection theory applies to integrals of the form of eq. (8). The first step is to endow the manifold with a variable ordering $\{z_{i_1}, \ldots, z_{i_n}\}$ where the $i_n$-index informs of the order of the variables from 1 to $n$. To calculate the multivariate intersection number for $n$-differential forms, it is necessary to know $\nu_\mathbf{k} = \dim(H^k)$. It can be obtained in a manner similar to the system of eq. (21) [23, 33]. The number $k$ of discrete elements is $\mathbf{k} \in \{i_1, \ldots, i_k\} \subset \{1, \ldots, n\}$. We obtain a set of dimensions $\{\nu_\mathbf{1}, \ldots, \nu_\mathbf{n}\}$ for the iterative integral in the variables $\{\{z_{i_1}\}, \ldots, \{z_{i_1}, \ldots, z_{i_n}\}\}$ respectively, such that $\mathbf{1} = \{i_1\}$, $\mathbf{2} = \{i_1, i_2\}$, ..., $\mathbf{n} = \{i_1, \ldots, i_2\}$. We must point out that although $\nu_\mathbf{n}$ is independent of the order of the integration variables, $\nu_\mathbf{k}$ may depend on which specific subset $\mathbf{k}$ of $\{1, 2, \ldots, n\}$ is chosen and in which order. As a working principle, we choose the order that minimizes $\nu_\mathbf{k}$ for all $\mathbf{k}$ forms $\mathbf{k} \in \{1, \ldots, n\}$. We can count the number of master integrals on the inner manifold, in a manner similar to that before; see eq. (21). For further discussion of the counting of variables in the individual layers, see e.g. ref. [30].

The aim is then to express the $n$-variable intersection number $\langle \varphi^{(\mathbf{n})} | \varphi^{(\mathbf{n}) \vee} \rangle$ in terms of the intersection number in $(n-1)$ variables on the inner manifold. The choice of variables and their orienting parametrization of the inner and outer manifolds is arbitrary. As before, we use $\mathbf{k} \equiv \{i_1, i_2, \ldots, i_k\}$ to denote the variables. Therefore the $\mathbf{n}$-forms $\langle \varphi^{(\mathbf{n})} | \in H^n(X, \nabla_\omega)$ and $\langle \varphi^{(\mathbf{n}), \vee} | \in H^n(X, \nabla_{-\omega})$ can be decomposed, and the decomposition formula takes the following form:

$$\left\langle \varphi^{(\mathbf{n})} \right| = \sum_{i=1}^{\nu_{\mathbf{n-1}}} \left\langle e_i^{(\mathbf{n-1})} \right| \wedge \left\langle \varphi_i^{(n)} \right| \tag{33}$$

$$\left| \varphi^{(\mathbf{n}) \vee} \right\rangle = \sum_{i=1}^{\nu_{\mathbf{n-1}}} \left| h_i^{(\mathbf{n-1})} \right\rangle \wedge \left| \varphi_i^{(n) \vee} \right\rangle \tag{34}$$

with an in principle arbitrary basis $(\langle e_1^{(\mathbf{n-1})}|, \ldots, \langle e_{\nu_{\mathbf{n-1}}}^{(\mathbf{n-1})}|)$ as the basis for $H^{n-1}(X_{n-1}, \nabla_{\omega | dz_n = 0})$ and likewise for its dual $(|h_1^{(\mathbf{n-1})}\rangle, \ldots, |h_{\nu_{\mathbf{n-1}}}^{(\mathbf{n-1})}\rangle)$. In the above expression, $\langle \varphi_i^{(n)} |$ and $|\varphi_i^{(n) \vee} \rangle$ are one-forms in the variables $z_n$ and are treated as coefficients of the basis expansion. They can be obtained by projection, using an analogue of the master decomposition formula eq. (24):

$$\left\langle \varphi_i^{(n)} \right| = \sum_{j=1}^{\nu_{\mathbf{n-1}}} \left\langle \varphi^{(\mathbf{n})} | h_j^{(\mathbf{n-1})} \right\rangle \left( \mathbf{C}_{(\mathbf{n-1})}^{-1} \right)_{ji} \tag{35}$$

$$\left| \varphi_i^{(n) \vee} \right\rangle = \sum_{j=1}^{\nu_{\mathbf{n-1}}} \left( \mathbf{C}_{(\mathbf{n-1})}^{-1} \right)_{ij} \left\langle e_j^{(\mathbf{n-1})} | \varphi^{(\mathbf{n}) \vee} \right\rangle \tag{36}$$

with

$$\left( \mathbf{C}_{(\mathbf{n-1})} \right)_{ij} = \left\langle e_i^{(\mathbf{n-1})} | h_j^{(\mathbf{n-1})} \right\rangle. \tag{37}$$

To define a generalization of the differential equation of eq. (29), we need to define a covariant connection on the whole manifold $X$, and for that we need to define a connection coefficient $\hat{\Omega}^n$ on $X$ that takes the nested structure of the manifold into account. That connection coefficient is a $\nu_{\mathbf{n-1}} \times \nu_{\mathbf{n-1}}$ matrix with entries given by

$$\hat{\Omega}_{ij}^{(n)} = \sum_{k=1}^{\nu_{(\mathbf{n-1})}} \left\langle \nabla_{\omega_{n-1}} e_i^{(\mathbf{n-1})} | h_k^{(\mathbf{n-1})} \right\rangle \left( \mathbf{C}_{(\mathbf{n-1})}^{-1} \right)_{kj} \tag{38}$$

The connection reads as follows:

$$(\nabla_{\Omega^{(n)}})_{ij} (\bullet)_j = \delta_{ij} d(\bullet)_j + (\Omega^{(n)})_{ij} \wedge (\bullet)_j \tag{39}$$

The differential equations then have the following form:

$$\nabla_{\Omega^n} \psi_{x_i}^{(\mathbf{n})} = \varphi^{(\mathbf{n})} \tag{40}$$

around each point $x_i$ from the set of $\mathcal{P}_\Omega^{(\mathbf{n})}$, where $\psi_{x_i}$ is a vector-valued solution of the differential equation eq. (40). As the connection coefficient $\Omega^{(n)}$ is matrix valued, the regularization map (corresponding to eq. (26) in the univariate case) now has the form

$$\mathcal{R}_{\Omega^{(n)}}(\varphi^{(n)}) = \varphi^{(n)} - \sum_{x_i \in \mathcal{P}_{\Omega^{(n)}}} \nabla_{\Omega^{(n)}} \left( h_{x_i}(z_n, \bar{z}_n) \psi_{x_i} \right) \tag{41}$$

The recursive formula for the intersection number then becomes

$$\left\langle \varphi^{(\mathbf{n})} | \varphi^{(\mathbf{n}) \vee} \right\rangle = \sum_{x_i \in \mathcal{P}_{\Omega^{(n)}}} \text{Res}_{z_n = x_i} \left( \psi_{x_i, i}^{(n)} \cdot \mathbf{C}_{ij}^{(\mathbf{n-1})} \cdot \varphi_j^{(n) \vee} \right) \tag{42}$$

This is the expression we will use for the computation of multivariate intersection numbers.

### E. Delta-bases and relative cohomology

There is a problem with the theory as discussed so far, which is that Feynman integrals in the Baikov representation as given by eq. (7) are not directly on the form of eq. (8). The reason for that is that the propagator powers $a_i$ of eq. (7) are integer, while the powers $\alpha_i$ of eq. (8) specifically have to be noninteger for the theory to be valid. One type of fix is an introduction of *regulators* $\rho$, in which we specifically introduce the noninteger powers as

$$u \to u_{\text{reg}} = u \prod_i z_i^{\rho_i} \quad (43)$$

and then at the very end of the calculation the regulators may be put to zero. While that approach is completely valid and produces correct results, also for the calculation considered in this paper [33], it introduces a new scale $\rho_i$ for each variable that significantly complicates the computations. We will therefore pursue a different approach, which is the introduction of *delta-bases*. The delta-bases as discussed here were introduced in ref. [28, 30, 48], and some of the mathematical background is given in ref. [53]. They have their origin in the mathematical framework of *relative cohomology* in which one explicitly works modulo the zero locus of the set of propagators going on-shell. The way this works in practice is that the new type of dual forms are allowed, which may be written as

$$\varphi^\vee = \delta_s \, \xi \quad (44)$$

where $s$ is some subset of the $n$ variables and $\xi$ is a $(n-|s|)$-form. Having such a delta in the dual basis makes the intersection number

$$\langle \phi | \delta_s \, \xi \rangle_n = \langle \text{Res}_{s_1=0,\ldots}(\phi) | \xi \rangle_{n-|s|} \quad (45)$$

In other words the $\delta$ works as a residue operator.

The benefits of the delta bases are threefold: First, they remove the need for the $\rho$-regulators discussed above. Second, they simplify the computation of the intersection numbers, in that eq. (42) now has to be applied, not for all the $n$ variables but only for $n-|s|$ whenever a delta is present. Finally, many of the intersection numbers will trivially be zero whenever the residue in eq. (45) vanishes, making the matrix $\mathbf{C}$ of eq. (23) block triangular.

We are now ready to apply our theory to the gravity problem under discussion.

## IV. PERFORMING THE REDUCTIONS

Looking at diagrams in fig. 2 we notice that they all would contribute to a *unitarity cut* in the $t$-channel. This means that we would not lose a single contribution by performing the double-cut of those two propagators. Doing so allows us to significantly simplify the computation, in a way that the intersection-based approach can take full advantage of, since all intersection numbers needed will be between two-forms as opposed to the four-forms that would be needed for the complete problem. Given the recursive nature of the multivariate algorithm, the step in complexity from two to four variables is very significant.

Considering first the box diagram (i.e. the first diagram in fig. 2), we may perform the Baikov parametrization and do the cut, yielding (up to an overall constant)

$$u = \mathcal{B}^{(d-5)/2} \quad (46)$$

where

$$\begin{aligned} \mathcal{B} = {} & 4m_2^2 z_1^2 - m_2^4 t - t(m_1^2 + z_1 - s)^2 - 4(m_1^2 + m_2^2 - s)z_1 z_2 \\ & + (4m_1^2 - t)z_2^2 + 2t\big((m_1^2 + s + z_1)(m_2^2 + z_2) - m_2^2 z_2\big) \end{aligned} \quad (47)$$

and where $z_1$ and $z_2$ are the Baikov variables corresponding to the two uncut propagators; in the parametrization of eq. (5) $z_1 = (k+p_1)^2 - m_1^2$ and $z_2 = (k-p_4)^2 - m_2^2$.

For the last four Feynman diagrams in fig. 2, the propagators are a subset of the propagators of the box, meaning that this representation is valid for them as well. For the crossed box (i.e. the second diagram in fig. 2) the parametrization is related to that of eq. (47) through $s \leftrightarrow u$, so we will not need to consider it separately.

Applying the Feynman rules as discussed in section II, we may then express all our Feynman integrals in terms of integrals over the variables $z_1$ and $z_2$. In particular, the following monomials appear:

$$\left\{ \tfrac{1}{z_1 z_2}, \tfrac{1}{z_1}, \tfrac{z_2}{z_1}, \tfrac{z_2^2}{z_1}, \tfrac{1}{z_2}, \tfrac{z_1}{z_2}, \tfrac{z_1^2}{z_2}, 1, z_1, z_1^2, z_2, z_2^2 \right\} \quad (48)$$

of which we would like to express the integrals in terms of a minimal basis of master integrals.

### A. Finding a basis

To determine the variable ordering for each layer of the fibration that gives the minimum basis of the cut-box and cross-box, we solve eq. (21) for the set of integration variables and their possible subsets, and get the dimensions of the twisted cohomology groups on each layer of the fibration.

| Solve | $z_1$ | $z_2$ | $z_1, z_2$ |
|---|---|---|---|
| $\nu$ | 2 | 2 | 4 |

We choose the variable ordering from the inside out $z_1, z_2$. This corresponds to there being two "master integrals" in the inner basis and four in the outer. The opposite ordering would have given the same counting, unsurprisingly given the symmetry of the problem.

We now want to figure out what choice of master integrals is valid in the different layers. This may be done by introducing regulators in the sense of eq. (43) and performing the counting in the presence and absence of each regulator, respectively. See ref. [30] for further details of this algorithm. The results are

| Regulated variables | Solve $\{z_1\}$: Basis size | Solve $\{z_1, z_2\}$: Basis size |
|---|---|---|
| None | 1 | 1 |
| $z_1$ | 2 | 2 |
| $z_2$ | 1 | 2 |
| $z_1, z_2$ | 2 | 4 |

From the middle column we conclude that if $z_1$ is not allowed to appear as a propagator in the inner basis, there will be one basis element; otherwise, there will be two. Thus a valid inner basis is

$$e^{(1)} = \left\{1, \frac{1}{z_1}\right\} dz_1 \qquad (49)$$

Using the same argument we likewise conclude for the outer basis that a valid set is

$$e^{(2)} = \left\{1, \frac{1}{z_1}, \frac{1}{z_2}, \frac{1}{z_1 z_2}\right\} dz_1 \wedge dz_2 \qquad (50)$$

This corresponds, of course, to the well-known set of master integrals consisting of a box, two triangles, and a bubble.

For the dual basis the prescription for delta-bases is that a pole in the basis translates to an index on the delta in the dual basis. Thus we get

$$h^{(1)} = \{1, \delta_{z_1}\}, \quad h^{(2)} = \{1, \delta_{z_1}, \delta_{z_2}, \delta_{z_1 z_2}\}. \qquad (51)$$

We are now ready to perform the reduction.

### B. Obtaining the Coefficients

To compute coefficients of the master integral bases for a given integral, via intersection theory, we use eq. (24). To obtain $\mathbf{C}_{ij} = \langle e_i | h_j \rangle$ and $\langle \varphi | h_j \rangle$ via the multivariate intersection number, we need to first decompose the integral in the Baikov representation into the regulated Baikov polynomial $u_{\text{reg}}(\mathbf{z})$ and the differential form $\varphi$. Then, we obtain the vectorial connection coefficient of the individual fibration $\omega$; see eq. (16). Furthermore, we need to know the basis of the individual layers of the fibration as well as their respective duals; see section IV A. The first step in the recursive multivariate intersection algorithm is to compute the innermost intersection number based solely on the inner basis and its dual via the univariate intersection number; see section III C.

Let us introduce the object

$$\begin{aligned}\lambda &= \Lambda(m_1^2, m_2^2, s) \\ &= m_1^4 + m_2^4 + s^2 - 2m_1^2 m_2^2 - 2m_1^2 s - 2m_2^2 s\end{aligned} \qquad (52)$$

where $\Lambda$ is the Källén function. This is related to the $\mathcal{B}$ of eq. (47) through

$$\mathcal{B}(z_1{=}0, z_2{=}0) = -t\lambda \qquad (53)$$

In terms of that object the inner intersection yields

$$\mathbf{C}^{(1)} = \begin{bmatrix} \frac{4(d-5)(\lambda+st)(m_2^2 t + t z_2 + z_2^2)}{(d-6)(d-4)(4m_2^2 - t)^2} & 0 \\ \frac{(m_1^2 - m_2^2 - s - z_2)t + 2(m_1^2 + m_2^2 - s)z_2}{(d-6)(4m_2^2 - t)} & 1 \end{bmatrix} \qquad (54)$$

and the connection matrix of eq. (38) becomes

$$\mathbf{\Omega}^T = \begin{bmatrix} \frac{(d-4)(t+2z_2)}{2(m_2^2 + t z_2 + z_2^2)} & \frac{-(d-4)(4m_2^2 - t)t(m_1^2 + m_2^2 - s + z_2)}{2(m_2^2 + t z_2 + z_2^2)\mathcal{B}(z_1 = 0)} \\ 0 & \frac{(d-5)(4m_1^2 z_2 - t(m_2^2 - m_1^2 - s + z_2))}{\mathcal{B}(z_1 = 0)} \end{bmatrix} \qquad (55)$$

We also need the pairings between the inner and the outer bases $\langle e^{(1)} | h^{(2)} \rangle$ and $\langle e^{(2)} | h^{(1)} \rangle$. We will not write those expressions here, see our ancillary file[*] for the full list.

We are then ready to perform the computation in the outer layer. We follow the algorithm outlined in section III D to obtain the quantities in eq. (24).

$$\mathbf{C} = \begin{bmatrix} C_{11} & 0 & 0 & 0 \\ C_{21} & C_{22} & 0 & 0 \\ C_{31} & 0 & C_{33} & 0 \\ C_{41} & C_{42} & C_{43} & 1 \end{bmatrix} \qquad (56)$$

where

$$\begin{aligned}C_{11} &= \frac{-t(\lambda + st)}{4(d-7)(d-3)} \\ C_{21} &= \frac{t\big((d-6)t - 4(2d-11)m_1^2\big)(\lambda + st)}{2(d-7)(d-6)(d-4)(4m_1^2 - t)^2} \\ C_{42} &= \frac{(m_2^2 - m_1^2 - s)t}{(d-6)(4m_1^2 - t)} \\ C_{41} &= \frac{t^2 \big(2(m_1^2 - m_2^2)^2 - 2(m_1^2 + m_2^2)s + st\big)}{(d-7)(d-6)(4m_1^2 - t)(4m_2^2 - t)} \\ C_{22} &= \frac{4(d-5)m_1^2(\lambda + st)}{(d-6)(d-4)(4m_1^2 - t)^2}\end{aligned} \qquad (57)$$

and where furthermore $\{C_{31}, C_{33}, C_{43}\}$ are related to $\{C_{21}, C_{22}, C_{42}\}$ through $m_1 \leftrightarrow m_2$. We see that $\mathbf{C}$ is block triangular as promised in section III E.

To perform the reductions of all the monomials of eq. (48), we also need the pairings of each of them with the dual basis $|h\rangle$ in accordance with eq. (24), which is the list of intersection numbers $\langle \varphi_i | h_j \rangle$. We will not list them here, see our ancillary file for the full list. Using eq. (24) we may then finally decompose the integrals of the monomials of eq. (48) onto our basis of master integrals. The results for those decompositions may likewise be found in our ancillary files. The results are found to be in agreement with the public IBP code FIRE [17].

Putting it all together we get four master integral coefficients, $c_1 - c_4$, for each of the two families (those of the box and the crossed box). We did this for each family, and the $s \leftrightarrow u$ symmetry between the two families may be used as a check. The triangles and the bubbles may be identified between the two integral families, so at the end we have five master integral coefficients

$$\{c_{\Box,s},\ c_{\Box,u},\ c_\triangle,\ c_\triangledown,\ c_{\text{bub}}\} \quad (58)$$

the results for which may also be found in our ancillary file.

---

[*] This file may be found on GitHub at *https://github.com/HjalteFrellesvig/GRfromInterX*

## V. THE MASTER INTEGRALS

In this section we will look at the master integrals that end up contributing to 2PM scattering. The section follows refs. [10, 54]. As mentioned in the introduction, an integral part of the PM expansion is the soft limit, in which the momentum transfer between the two compact objects is much smaller than other scales in the problem, i.e. $|t| \ll s, m_1^2, m_2^2$. This may be approached with the *method of regions* [55, 56] in which we take the loop momenta of the integrals to scale with the size of the transferred momentum, extracting the *soft region*, i.e. $k \sim q$ where $q = p_1+p_2$. To take the classical limit we will also want to introduce a classical momentum transfer as $\tilde{q} = q/\hbar$. We may then express our master integrals in terms of a series expansion in $\tilde{q}$. Looking first at the box contributions, it turns out that the box $\mathcal{I}_{\Box,s}$ and the crossed box $\mathcal{I}_{\Box,u}$ only appear through their sum due to the respective coefficients becoming the same in the $t \to 0$ limit. This induces some simplifications, and following refs. [10, 33, 54] the result may be written

$$\begin{aligned}\mathcal{I}_{\Box,\Sigma} &= \mathcal{I}_{\Box,s} + \mathcal{I}_{\Box,u} \\ &= (\tilde{q}^2)^{-\epsilon}\left(\mathcal{I}_\Box^{(-1)}\frac{1}{|\tilde{q}|} + \mathcal{I}_\Box^{(0)} + \mathcal{O}(|\tilde{q}|)\right)\end{aligned} \quad (59)$$

where

$$\mathcal{I}_\Box^{(-1)} = \frac{\Gamma(-\epsilon)^2\Gamma(1-\epsilon) - \pi}{2\Gamma(-2\epsilon)\sqrt{(p_1\cdot p_4)^2 - m_1^2 m_2^2}} \quad (60)$$

$$\mathcal{I}_\Box^{(0)} = \frac{\Gamma(\tfrac{1}{2}+\epsilon)\Gamma(\tfrac{1}{2}-\epsilon)^2}{2\Gamma(-2\epsilon)}\frac{m_1+m_2}{(p_1\cdot p_4)^2 - m_1^2 m_2^2} \quad (61)$$

where we have introduced $\epsilon = 2 - d/2$. A similar computation for the triangles gives

$$\mathcal{I}_\triangledown = (\tilde{q}^2)^{-\epsilon}\left(\frac{\Gamma(\tfrac{1}{2}+\epsilon)\Gamma(\tfrac{1}{2}-\epsilon)^2}{m_1\Gamma(1-2\epsilon)} + \mathcal{O}(|\tilde{q}|)\right) \quad (62)$$

and likewise for $\mathcal{I}_\triangle$ with $m_1 \leftrightarrow m_2$. Finally, the bubble integral turns out to vanish in the soft limit and will not concern us further.

One might worry about the doubly divergent, both in $|q|$ and in $\epsilon$, contribution $\mathcal{I}_\Box^{(-1)}$. Yet it turns out that when computing a finite physical quantity, such as a scattering angle, this divergence cancels with a divergence arising from the square of the 1PM contribution, in a manner similar to the cancellation of infrared divergences in standard QFTs.

## VI. RESULTS

We are now prepared to write down an expression for our results. We may express our 2PM contribution as

$$\mathcal{M}_{\text{2PM}} = \kappa^4\left(c_\Box \mathcal{I}_{\Box,\Sigma} + c_\triangledown \mathcal{I}_\triangledown + c_\triangle \mathcal{I}_\triangle\right) \quad (63)$$

where

$$c_\Box = \frac{\left((d-2)(m_1^2+m_2^2-s)^2 - 4m_1^2 m_2^2\right)^2}{16(d-2)^2}, \quad (64)$$

$$c_\triangle = \frac{-m_2^2\left((\lambda+2m_1^2 m_2^2)(4d^2-17d+19) + 2m_1^2 m_2^2(4d^2-15d+5)\right)}{16(d-2)^2} \quad (65)$$

where $\lambda$ is given by eq. (52). We have used that $c_{\Box;s} = c_{\Box;u} = c_\Box$, and we also use that $c_\triangledown$ is related to $c_\triangle$ through $m_1 \leftrightarrow m_2$. These coefficients equal those of eq. (58) in the $t \to 0$ limit.

In particular, if we insert $d=4$ these coefficients reduce to

$$\begin{aligned}c_\Box &= \tfrac{1}{4}\left(m_1^2 m_2^2 - 2(p_1\cdot p_4)^2\right)^2 \\ c_\triangle &= \tfrac{3}{16}m_2^2\left(m_1^2 m_2^2 - 5(p_1\cdot p_4)^2\right) \\ c_\triangledown &= \tfrac{3}{16}m_1^2\left(m_1^2 m_2^2 - 5(p_1\cdot p_4)^2\right)\end{aligned} \quad (66)$$

in agreement with the expressions given in refs. [6, 57].

## VII. DISCUSSION AND OUTLOOK

In section II we introduced the post-Minkowskian approach to classical gravity. We also introduced the Baikov representation, and showed the diagrams contributing to the 2PM contribution.

In section III, which forms the bulk of the paper, we introduced the intersection theory that we use for the integral decomposition: In sec. III A we introduced some concepts from twisted cohomology theory, which is the mathematical framework we use. In sec. III B we showed how the existence of an inner product on the cohomology group allows for a direct path to integral decomposition using the master decomposition formula eq. (24). In sec. III C we showed how to define that inner product, the intersection number, for the case of univariate integrals. In sec. III D we introduce the multivariate intersection number in a fibration-based approach, including the main formula used to compute them eq. (42). Finally, in sec. III E we introduce delta-bases that are needed to compute intersection numbers of objects involving unregulated poles corresponding to uncut propagators.

In section IV we performed the reduction of the Feynman integrals needed for the 2PM computation onto a basis that we identified through a systematic analysis. The intermediate and final results may be found in our ancillary file. In section V we showed the results for the master integrals as may be obtained using the soft limit in the expansion by regions. Finally in section VI we put it all together into a result for the 2PM contribution, which was found to be in agreement with the literature [6, 10, 54].

This provides evidence for the applicability of intersection theory to the post-Minkowskian expansion.

It has been proposed in ref. [58] (under the name of the *bottom-up approach*) to apply the intersection-based master integral decomposition on a set of *spanning cuts*, of which each is simpler that performing the full reduction. The spanning cuts are defined such that each master integral contributes to at least one cut. Yet we see that due to the nature of the soft limit taken in the PM approach, all the diagrams contributing at 2PM could be extracted from a single $t$-channel cut. In other words, the set of spanning cuts has only one member at 2PM! This property will generalize to higher orders in the PM expansions, where the set of spanning cuts still only contains cuts in the $t$-channel even though there will be several of those at higher orders. Doing the 3PM calculation with the methods outlined here would be within reach.

In the work in this paper we performed the reduction onto the master integral basis before taking the soft limit. Yet many approaches reverse the order of the two and consider the integrals in the soft limit only, something that significantly simplifies the computations, linearizes the matter propagators $((k-p)^2 - m^2 \to 2k \cdot v)$, and makes the integrals effectively one-scale objects [8, 9]. We decided against that approach in this paper, but there is nothing preventing the use of intersection-based methods to these soft limit integrals, and presumably many of our intermediate expressions would simplify with such an approach. For the use of the Baikov representation on such linearized integrals see e.g. ref. [59].

The recent years have seen an explosive growth in the number of works applying intersection theory in various branches of physics. Applications have been found in fields as diverse as cosmology [60, 61], lattice gauge theory [62, 63], and gravitational phase-space integrals [64], and this first application in the context of the post-Minkowskian expansion of general relativity adds another branch to this growth.

We hope with this work to have added a useful new perspective to the ongoing investigations into gravitational waves and black hole physics, further expanding the large set of amplitude methods that are found to be applicable in this new regime.

### Acknowledgements

We wish to thank Poul Henrik Damgaard for many insightful discussions during this project. We also thank Poul Henrik Damgaard, Roger Morales, Sid Smith, and Giacomo Brunello for reading through the manuscript in its draft stages and providing thorough comments.

H.F. is supported by the research grant no. 00025445 from Villum Fonden, by a Carlsberg Foundation Reintegration Fellowship, and by the European Union's Horizon 2020 research and innovation programme under the Marie Skłodowska-Curie grant agreement No. 847523 "INTERACTIONS".

[1] B. P. Abbott *et al.* (LIGO Scientific, Virgo), Phys. Rev. Lett. **116**, 061102 (2016), arXiv:1602.03837 [gr-qc].
[2] B. P. Abbott *et al.* (LIGO Scientific, Virgo), Phys. Rev. Lett. **119**, 161101 (2017), arXiv:1710.05832 [gr-qc].
[3] T. Damour, Phys. Rev. D **94**, 104015 (2016), arXiv:1609.00354 [gr-qc].
[4] A. Buonanno, M. Khalil, D. O'Connell, R. Roiban, M. P. Solon, and M. Zeng, in *Snowmass 2021* (2022) arXiv:2204.05194 [hep-th].
[5] N. E. J. Bjerrum-Bohr, P. H. Damgaard, G. Festuccia, L. Planté, and P. Vanhove, Phys. Rev. Lett. **121**, 171601 (2018), arXiv:1806.04920 [hep-th].
[6] C. Cheung, I. Z. Rothstein, and M. P. Solon, Phys. Rev. Lett. **121**, 251101 (2018), arXiv:1808.02489 [hep-th].
[7] D. A. Kosower, B. Maybee, and D. O'Connell, JHEP **02**, 137 (2019), arXiv:1811.10950 [hep-th].
[8] Z. Bern, C. Cheung, R. Roiban, C.-H. Shen, M. P. Solon, and M. Zeng, Phys. Rev. Lett. **122**, 201603 (2019), arXiv:1901.04424 [hep-th].
[9] Z. Bern, C. Cheung, R. Roiban, C.-H. Shen, M. P. Solon, and M. Zeng, JHEP **10**, 206 (2019), arXiv:1908.01493 [hep-th].
[10] A. Cristofoli, P. H. Damgaard, P. Di Vecchia, and C. Heissenberg, JHEP **07**, 122 (2020), arXiv:2003.10274 [hep-th].
[11] G. Kälin and R. A. Porto, JHEP **11**, 106 (2020), arXiv:2006.01184 [hep-th].
[12] G. Mogull, J. Plefka, and J. Steinhoff, JHEP **02**, 048 (2021), arXiv:2010.02865 [hep-th].
[13] N. E. J. Bjerrum-Bohr, P. H. Damgaard, L. Plante, and P. Vanhove, J. Phys. A **55**, 443014 (2022), arXiv:2203.13024 [hep-th].
[14] G. Kälin and R. A. Porto, JHEP **01**, 072 (2020), arXiv:1910.03008 [hep-th].
[15] K. G. Chetyrkin and F. V. Tkachov, Nucl. Phys. B **192**, 159 (1981).
[16] S. Laporta, Int. J. Mod. Phys. A **15**, 5087 (2000), arXiv:hep-ph/0102033.
[17] A. V. Smirnov and F. S. Chuharev, Comput. Phys. Commun. **247**, 106877 (2020), arXiv:1901.07808 [hep-ph].
[18] J. Klappert, F. Lange, P. Maierhöfer, and J. Usovitsch, Comput. Phys. Commun. **266**, 108024 (2021), arXiv:2008.06494 [hep-ph].
[19] K. Cho and K. Matsumoto, Nagoya Math. J. **139**, 67 (1995).
[20] K. Matsumoto, Osaka J. Math. **35**, 873 (1998).
[21] S. Mizera, Phys. Rev. Lett. **120**, 141602 (2018), arXiv:1711.00469 [hep-th].
[22] P. Mastrolia and S. Mizera, JHEP **02**, 139 (2019), arXiv:1810.03818 [hep-th].

[23] H. Frellesvig, F. Gasparotto, S. Laporta, M. K. Mandal, P. Mastrolia, L. Mattiazzi, and S. Mizera, JHEP **05**, 153 (2019), arXiv:1901.11510 [hep-ph].
[24] H. Frellesvig, F. Gasparotto, M. K. Mandal, P. Mastrolia, L. Mattiazzi, and S. Mizera, Phys. Rev. Lett. **123**, 201602 (2019), arXiv:1907.02000 [hep-th].
[25] S. Mizera and A. Pokraka, JHEP **02**, 159 (2020), arXiv:1910.11852 [hep-th].
[26] J. Chen, X. Jiang, X. Xu, and L. L. Yang, Phys. Lett. B **814**, 136085 (2021), arXiv:2008.03045 [hep-th].
[27] S. Weinzierl, J. Math. Phys. **62**, 072301 (2021), arXiv:2002.01930 [math-ph].
[28] S. Caron-Huot and A. Pokraka, JHEP **12**, 045 (2021), arXiv:2104.06898 [hep-th].
[29] G. Fontana and T. Peraro, JHEP **08**, 175 (2023), arXiv:2304.14336 [hep-ph].
[30] G. Brunello, V. Chestnov, G. Crisanti, H. Frellesvig, M. K. Mandal, and P. Mastrolia, (2023), arXiv:2401.01897 [hep-th].
[31] S. Weinzierl, *Feynman Integrals* (Springer Cham, 2022) arXiv:2201.03593 [hep-th].
[32] F. Gasparotto, *Co-Homology and Intersection Theory for Feynman Integrals*, Ph.D. thesis, U. Padua (main) (2023).
[33] T. Teschke, *General Relativity from Intersection Theory and Loop Integrals*, Master's thesis, Bohr Inst. (2024), arXiv:2401.01920 [hep-th].
[34] Z. Bern, J. Parra-Martinez, R. Roiban, M. S. Ruf, C.-H. Shen, M. P. Solon and M. Zeng, Phys. Rev. Lett. **126**, 171601 (2021), arXiv:2101.07254 [hep-th].
[35] C. Dlapa, G. Kälin, Z. Liu, and R. A. Porto, Phys. Lett. B **831**, 137203 (2022), arXiv:2106.08276 [hep-th].
[36] G. U. Jakobsen, G. Mogull, J. Plefka, B. Sauer, and Y. Xu, Phys. Rev. Lett. **131**, 151401 (2023), arXiv:2306.01714 [hep-th].
[37] H. Frellesvig, R. Morales, and M. Wilhelm, Phys. Rev. Lett. **132**, 201602 (2024), arXiv:2312.11371 [hep-th].
[38] A. Klemm, C. Nega, B. Sauer, and J. Plefka, Phys. Rev. D **109**, 124046 (2024), arXiv:2401.07899 [hep-th].
[39] M. Driesse, G. U. Jakobsen, G. Mogull, J. Plefka, B. Sauer, and J. Usovitsch, Phys. Rev. Lett. **132**, 241402 (2024), arXiv:2403.07781 [hep-th].
[40] G. U. Jakobsen, *General Relativity from Quantum Field Theory*, Master's thesis, Bohr Inst. (2020), arXiv:2010.08839 [hep-th].
[41] G. U. Jakobsen, Phys. Rev. D **102**, 104065 (2020), arXiv:2006.01734 [hep-th].
[42] D. Neill and I. Z. Rothstein, Nucl. Phys. B **877**, 177 (2013), arXiv:1304.7263 [hep-th].
[43] P. A. Baikov, Nucl. Instrum. Meth. A **389**, 347 (1997), arXiv:hep-ph/9611449.
[44] H. Frellesvig and C. G. Papadopoulos, JHEP **04**, 083 (2017), arXiv:1701.07356 [hep-ph].
[45] X. Jiang and L. L. Yang, Phys. Rev. D **108**, 076004 (2023), arXiv:2303.11657 [hep-ph].
[46] K. Aomoto and M. Kita, *Theory of Hypergeometric Functions* (Springer, Tiergartenstraße 15 – 17 69121 Heidelberg, Germany, 2011).
[47] H. A. Frellesvig and L. Mattiazzi, PoS **MA2019**, 017 (2022), arXiv:2102.01576 [hep-ph].
[48] S. Caron-Huot and A. Pokraka, JHEP **04**, 078 (2022), arXiv:2112.00055 [hep-th].
[49] M. Giroux and A. Pokraka, JHEP **03**, 155 (2023), arXiv:2210.09898 [hep-th].
[50] R. N. Lee and A. A. Pomeransky, JHEP **11**, 165 (2013), arXiv:1308.6676 [hep-ph].
[51] S. Mizera, *Aspects of Scattering Amplitudes and Moduli Space Localization*, Ph.D. thesis, Princeton, Inst. Advanced Study (2020), arXiv:1906.02099 [hep-th].
[52] V. Chestnov, H. Frellesvig, F. Gasparotto, M. K. Mandal, and P. Mastrolia, JHEP **06**, 131 (2023), arXiv:2209.01997 [hep-th].
[53] K. Matsumoto, "Relative twisted homology and cohomology groups associated with lauricella's $f_d$," (2019), arXiv:1804.00366 [math.AG].
[54] A. Koemans Collado, P. Di Vecchia, and R. Russo, Phys. Rev. D **100**, 066028 (2019), arXiv:1904.02667 [hep-th].
[55] M. Beneke and V. A. Smirnov, Nucl. Phys. B **522**, 321 (1998), arXiv:hep-ph/9711391.
[56] V. A. Smirnov, Springer Tracts Mod. Phys. **177**, 1 (2002).
[57] A. Cristofoli, N. E. J. Bjerrum-Bohr, P. H. Damgaard, and P. Vanhove, Phys. Rev. D **100**, 084040 (2019), arXiv:1906.01579 [hep-th].
[58] H. Frellesvig, F. Gasparotto, S. Laporta, M. K. Mandal, P. Mastrolia, L. Mattiazzi, and S. Mizera, JHEP **03**, 027 (2021), arXiv:2008.04823 [hep-th].
[59] H. Frellesvig, R. Morales, and M. Wilhelm, (2024), arXiv:2405.17255 [hep-th].
[60] S. De and A. Pokraka, JHEP **03**, 156 (2024), arXiv:2308.03753 [hep-th].
[61] G. Brunello, G. Crisanti, M. Giroux, P. Mastrolia, and S. Smith, Phys. Rev. D **109**, 094047 (2024), arXiv:2311.14432 [hep-th].
[62] S. Weinzierl, Phys. Lett. B **805**, 135449 (2020), arXiv:2003.05839 [hep-th].
[63] F. Gasparotto, A. Rapakoulias, and S. Weinzierl, Phys. Rev. D **107**, 014502 (2023), arXiv:2210.16052 [hep-th].
[64] G. Brunello and S. De Angelis, JHEP **07**, 062 (2024), arXiv:2403.08009 [hep-th].